\documentclass[iop,apj]{emulateapj}
\slugcomment{{\sc Accepted to ApJ:} May 27, 2011}
\newcounter{subfigure}

\begin{document}

\title{A Population of Accreted SMC Stars in the LMC}

\author{Knut A.G. Olsen\altaffilmark{1}\altaffilmark{2}}
\affil{National Optical Astronomy Observatory, Tucson, AZ 85719}
\email{kolsen@noao.edu}
\altaffiltext{1}{Visiting Astronomer, Cerro Tololo Inter-American Observatory.
CTIO is operated by AURA, Inc.\ under contract to the National Science
Foundation.}
\altaffiltext{2}{This work is based in part on observations made with the Spitzer Space Telescope, which is operated by the Jet Propulsion Laboratory, California Institute of Technology under a contract with NASA.}
\author{Dennis Zaritsky}
\affil{Steward Observatory, University of Arizona, 933 North Cherry Avenue, Tucson, AZ 85721}
\email{dzaritsky@as.arizona.edu}
\author{Robert D. Blum}
\affil{National Optical Astronomy Observatory, Tucson, AZ 85719}
\email{rblum@noao.edu}
\author{Martha L. Boyer}
\affil{Space Telescope Science Institute, 3700 San Martin Drive, Baltimore, MD 21218}
\email{mboyer@stsci.edu}
\author{Karl D. Gordon}
\affil{Space Telescope Science Institute, 3700 San Martin Drive, Baltimore, MD 21218}
\email{kgordon@stsci.edu}

\begin{abstract}
We present an analysis of the stellar kinematics of the Large Magellanic Cloud based on $\sim$5900 new and existing velocities of massive red supergiants, oxygen-rich and carbon-rich AGB stars, and other giants.  After correcting the line-of-sight velocities for the LMC's space motion and accounting for asymmetric drift in the AGB population, we derive a rotation curve that is consistent with all of the tracers used, as well as that of published \ion{H}{1} data.  The amplitude of the rotation curve is $v_0=87\pm5$ km $s^{-1}$ beyond a radius $R_0=2.4\pm0.1$ kpc, and has a position angle of the kinematic line of nodes of $\theta=142\arcdeg\pm5\arcdeg$.  
By examining the outliers from our fits, we identify a population of 376 stars, or $\gtrsim$5\% of our sample, that have line-of-sight velocities that apparently oppose the sense of rotation of the LMC disk.  We find that these kinematically distinct stars are either counter-rotating in a plane closely aligned with the LMC disk, or rotating in the same sense as the LMC disk, but in a plane that is inclined by 54$\arcdeg\pm2\arcdeg$ to the LMC.
Their kinematics clearly link them to two known \ion{H}{1} arms, which have previously been interpreted as being pulled out from the LMC.  We measure metallicities from the Ca triplet lines of $\sim$1000 LMC field stars and 30 stars in the kinematically distinct population.  For the LMC field, we find a median [Fe/H]=$-0.56\pm0.02$ with dispersion of 0.5 dex, while for the kinematically distinct stars the median [Fe/H] is $-1.25\pm0.13$ with a dispersion of 0.7 dex.  The metallicity differences provide strong evidence that the kinematically distinct population originated in the SMC.  This interpretation has the consequence that the \ion{H}{1} arms kinematically associated with the stars are likely falling into the LMC, instead of being pulled out.

\end{abstract}

\section{Introduction}
The Magellanic Clouds have always been full of surprises, many of which stem from the fact that the Clouds are in the midst of interacting with each other and with the Milky Way.  A famous surprise was the discovery of the Magellanic Stream trailing the Clouds (Wannier \& Wrixon 1972; Mathewson et al.\ 1974), which is now known to have a companion Leading Arm (Putman et al.\ 1998), and which taken together extend for $\sim$200$\arcdeg$ across the sky (Nidever et al.\ 2010).  Other striking evidence of the interaction is found in the Small Magellanic Cloud's and the Large Magellanic Cloud's intrinsically elongated shapes (Caldwell \& Coulson 1986, van der Marel 2001), the possible warps of the LMC (Olsen \& Salyk 2002, Nikolaev et al.\ 2004, Subramaniam \& Subramanian 2009), and the young population of stars formed from tidal debris in the Magellanic Bridge (Harris 2007).  A recent surprise is that with new proper motion measurements of the Clouds (Kallivayalil et al.\ 2006a, 2006b; Piatek et al.\ 2008), the trajectories of the Clouds indicate that they are likely on first or second approach to the Milky Way (Besla et al.\ 2007, Boylan-Kolchin et al.\ 2010), such that most of the interaction effects, e.g.\ the Stream, must be due to the LMC-SMC binary pair interaction, not an interaction with the Milky Way (Besla et al.\ 2010).  Many further surprises are likely in store, as foreshadowed by Mu\~{n}oz et al.\ (2006), who found LMC stars 20 kpc away in the foreground of the Carina dwarf, and Saha et al.\ (2010), who showed that the LMC maintains an exponential disk profile out to at least 12 disk scale lengths.

An observed feature found in some merged or interacting disk galaxies is the presence of extended counter-rotating stellar disks.  The first, remarkable discovery of such a system is NGC 4550, an S0 galaxy with two roughly equal mass counter-rotating stellar disks (Rubin et al.\ 1992, Rix et al.\ 1992).  The NGC 5719/13 system appears to be an example of an interacting system where a significant counter-rotating stellar disk is forming (Vergani et al.\ 2007).  However, Kuijken et al.\ (1996) found no additional counter-rotating cases in a sample of 28 S0s, and concluded that no more than $\sim$5\% of the stars in these galaxies could be counter-rotating, a testament to the difficulty of detecting such stars.  

We report on the first results from an extensive spectroscopic study of the LMC.  As described below, we have found that the LMC contains a significant ($\sim5$\%) stellar population whose kinematics indicate either counter-rotation in a plane closely aligned with the LMC disk, or rotation in a plane that is highly inclined to the LMC disk.  Evidence for counter-rotation in the core (radius$\lesssim$2.5 kpc) of the LMC was previously presented by Subramaniam \& Prabhu (2005), while Kunkel et al.\ (1997) presented evidence for an out-of-plane polar ring in the LMC. 
Here, we present clear evidence of such a kinematically distinct stellar population, as well as evidence that the population is kinematically cold and originated in the SMC.  The LMC thus appears to have accreted stars from the SMC, and may be a unique laboratory for studying the formation counter-rotating disks.

\section{Observations}
The primary observational data used in this paper was collected during
the nights 20 - 26 Nov 2007 at the CTIO 4-m Blanco telescope with the
Hydra-CTIO multi-fiber spectrograph \citep{barden98}, with some
additional data obtained on 19 Nov 2007 during time shared with
instrument checkout and engineering.  All of the nights were clear,
with only some thin cirrus on the night of 21 Nov.  We used Hydra-CTIO
with the 400mm Schmidt camera and SITe 2K$\times$4K CCD, giving a gain
of 0.84 e$^-$/ADU and 3 e$^-$ read noise.  We observed with the
red-blazed KPGLF grating at a central wavelength of 8000\AA ~and an
OG515 order-blocking filter.  We masked the fibers with a
100\micron\ slit mask, yielding a 2-pixel spectral resolution of
$R\sim7000$ and velocity precision of $\sim$2$-$3 km s$^{-1}$, over a
wavelength range of $\sim$6850--9150\AA.  In addition to our program
targets, each night we observed several spectral template stars from a
list roughly matching the spectral types of our targets; these template
objects consisted of the stars HD223311 (spectral type K4III),
HD16115 (R-type carbon star), HD80170 (K5 III), and BN Mon (N-type carbon star). 

Our targets for the Hydra-CTIO observations were selected using the
Spitzer SAGE survey of the LMC \citep{meixner06} in combination with
the near-infrared photometry from 2MASS \citep{skrutskie06} and
optical photometry from the Magellanic Clouds Photometric Survey
\citep{zaritsky04}.  We identified candidate LMC red supergiants
(RSGs; evolved from stars with initial masses $\gtrsim$10 M$_\odot$),
giants evolved from stars with initial masses $\sim$5 M$_\odot$,
carbon-rich AGB stars, oxygen-rich AGB stars, as well as ``extreme''
AGB stars.  The definitions of these classes were taken from the
analysis of the Spitzer SAGE photometry by \citet{Blum06} and by
matching the above photometric databases with the positions of
previously confirmed RSGs \citep{massey03} and carbon stars
\citep{hardy}.  We chose these target classes both because they span a
range of stellar ages and because their numerous spectral features
provide precise velocity measurements.  We also chose to observe
558 candidate bright LMC main sequence stars; nearly all of
these turned out to lie in the Milky Way foreground, however, and are
not discussed further.  We assigned the chosen targets to individual
Hydra-CTIO fibers using the program hydraassign (Massey 1995), which
resulted in 62 configurations with $\sim$95 targets per configuration.
All of the remaining $\sim$30 fibers were used to measure the sky.
Fig.\ 1 shows the $J-[3.6], [3.6]$ color-magnitude diagram (CMD) of stars for which we measured spectra, compared to all of the available sources.
In all, we obtained useful spectra of 4734 targets, which excludes the 558 main sequence stars in the Milky Way foreground that we observed.  4567, or 96\%, of these targets were found to have velocities consistent with LMC membership.

\section{Data Reduction and Analysis}
We processed our data using IRAF\footnote{IRAF is distributed by the
  National Optical Astronomy Observatory, which is operated by the
  Association of Universities for Research in Astronomy, Inc., under
  cooperative agreement with the National Science Foundation.}
routines.  In brief, we used CCDPROC to subtract the overscan region,
to subtract a bias frame, and to trim extraneous pixels from the image
edges.  We then used the routine L.A.Cosmic \citep{vandokkum01} to
remove cosmic rays from the individual spectral images.  We finally
used DOHYDRA to extract the individual fiber spectra, apply the dome
flat field, solve for and apply the dispersion solution based on our
He/Ne/Ar arc lamp observations, and subtract the average sky spectrum.

We measured velocities for the LMC stars using cross-correlation
\citep{tonry79} with our template spectra of stars with known radial
velocities, using the IRAF task FXCOR.  We used FXCOR to first apply a
ramp filter in the Fourier domain to dampen the highest frequency
(e.g. noise) and lowest frequency (e.g. continuum) Fourier components,
and then to fit the narrow cross-correlation peaks with Lorentzians to
find their centroids.  We performed the cross-correlation in
wavelength windows that avoid the strongest telluric absorption
features, and used the template with the strongest correlation peak,
as measured by the $r$ parameter, to make the velocity measurement.
We used the errors reported by FXCOR, which derive from $r$, as the
velocity error.  We accepted only those measurements with $r>5$, which
in practice translated to velocity errors $\lesssim$5 km s$^{-1}$; the
average velocity error was measured to be 2.2 km s$^{-1}$.  We
confirmed the stability of our velocity measurements by
cross-correlating template spectra taken at different times during
each night and on different nights.  In all cases, the template
velocities agreed to within 1$-$2 km s$^{-1}$, such that velocity drifts across nights should be no larger than the measured random errors of $\sim$2
km s$^{-1}$.

A possible concern with regard to the velocities is systematic error
introduced by the mismatch in the spectral types of our program stars
with that of our template stars. In particular, a number of our
program stars have M-type spectra, whereas our closest velocity
templates for these stars have mid-K-type spectra.  Despite the
visible differences between K giant and M supergiant spectral types
(most notably the strong TiO bands present in the M supergiants),
there are still many spectral lines common to both types, e.g.\ the Ca
II triplet lines at $\sim$8600 \AA.  In Fig.\ 2, we show
representative spectra of a carbon star, an oxygen-rich AGB star, a red supergiant, and a $\sim$5 M$_{\odot}$ giant compared
to the velocity template used for each star.  The spectra are not flux-calibrated.  The color of the carbon star suggests that it is R-type, whereas the template star shown, BN Mon, is N-type.  Judging from the prominent TiO absorption at $\sim$7100 \AA ~and $\sim$8400 \AA, the O-rich AGB star is a mid-M giant (Silva \& Cornell 1992), while the template shown is type K4III.  The presence of TiO absorption in the spectrum of the red supergiant suggests that is a late K supergiant, whereas the template is K5III.  Finally, the $\sim$5 M$_{\odot}$ giant appears to be a mid-K-type star, as is its template.
As seen in the figure,
we obtained strong, symmetric cross-correlation peaks for each of the
spectra, even in cases where our targets had later spectral types than
the templates.  In addition, after applying the computed velocity
shift, the 8662 Ca II line clearly falls at the same wavelength in the
template and the program star.  We thus do not think that template
mismatches are a significant source of systematic error in the
velocities.

As a final check, we compared our velocity measurements of stars that
overlapped with previous work.  For the 19 RSGs in common with Massey
\& Olsen (2003), we found an average difference of $v_{\rm hel}-v_{\rm hel,MO}$=$3.9\pm1.2$ km s$^{-1}$, where the error is the standard error of the mean, and a median
difference of 2.8 km s$^{-1}$.  For the 97 RSGs in common with
Pr\'{e}vot et al.\ (1985), who observed their sample with the precise
velocity scanner CORAVEL (Baranne et al.\ 1979), we found an average
difference of $v_{\rm hel}-v_{\rm hel,  Prevot}$=$0.6\pm0.5$ km s$^{-1}$.  For the 91 carbon stars in
common with Hardy et al.\ (2001), the average difference is $v_{\rm
  hel}-v_{\rm hel, Hardy}$=$-2.4\pm0.4$km s$^{-1}$.
The very close agreement with previous work reinforces our conclusion
that systematic errors in our velocities are small.

We performed our kinematic analysis on the 4567 LMC RSGs, giants, and AGB
stars observed with Hydra-CTIO, the 857 carbon stars observed by
\citet{kunkel97} and Hardy et al. (2001), and the 481 RSGs observed by
\citet{massey03} and \citet{prevot85}--a total of $\sim$5900 stars.
As noted in the comprehensive paper of van der Marel et al.\ (2002;
hereafter vdM02), interpretation of radial velocities in the LMC must
account both for contribution of the LMC's space motion projected into
the line of sight, as a function of position on the sky, and for the
possibility of precession and nutation of the LMC disk.  Indeed, as
shown by \citet{vandermarel02}, the space motion contribution has a
larger effect on the line-of-sight velocity distribution than do the
LMC's internal motions.  The accurate LMC proper measurement by
\citet{kallivayalil06a}, updated by \citet{piatek08}, fixes the space
motion contribution and thus greatly improves the accuracy of the
kinematic analysis.

As we did in Olsen \& Massey (2007, hereafter Paper I), we followed
the procedure outlined by vdM02 to solve for the LMC's systemic
velocity $v_{\rm sys}$, the rate of change of the disk's inclination
d$i$/d$t$, the position angle of the kinematic line of nodes $\theta$,
and the LMC's internal rotation curve $v(R)$.  We fixed the location
of the LMC's dynamical center to the position determined by vdM02,
$\alpha_{\rm CM}=5^{\rm h}27^{\rm m}36^{\rm s}$, $\delta_{\rm
  CM}=-69\arcdeg52\arcmin12\arcsec$, as this position coincides
closely with the LMC's geometric center as determined from structure
out to large radii (van der Marel \& Cioni 2001).  We furthermore set
the inclination of the disk to $i=34\fdg7$, as determined by studies
of the LMC's geometry (van der Marel \& Cioni 2001), and parameterized
the LMC's rotation curve by a linear function with a value of 0 km
s$^{-1}$ at radius $R=0$ and rising to a value $v_0$ at radius $R_0$,
with $v(R)=v_0$ for $R>R_0$ (as done by \citet{piatek08}).  We
calculated errors in the fitted quantities by creating 10000 Monte
Carlo samples of our data, refitting the parameters, and calculating
the dispersion in the parameters.  The Monte Carlo samples were
produced through a bootstrap process, in which we populated the
samples by randomly drawing stars from our original dataset, but
allowing any individual star to be drawn any number of times in the
same sample.

Fig.\ 3 (left) shows a representation of the result of our fit to the
738 RSG line-of-sight velocities, which yielded parameters of $v_{\rm
  sys}=263\pm2$ km s$^{-1}$, d$i$/d$t$=$-184\pm81\arcdeg$ Gyr$^{-1}$,
$\theta=142\pm5\arcdeg$, $v_0=87\pm5$ km s$^{-1}$, and $R_0=2.4\pm0.1$
kpc.  Following Paper I, in making this figure we first removed the
contribution of the LMC's space motion and the contribution from
d$i$/d$t$, then assumed that the remaining velocity signatures were
due to in-plane circular motions in the LMC disk.  For comparison, we
also show the emission-weighted mean \ion{H}{1} velocities from
\citet{kim98} and \citet{ss03}, with the same corrections applied as
to the RSGs.  Note that while the fit was performed on the
line-of-sight velocities, i.e.\ the projected velocities due to the
LMC's internal kinematics and space motion, in this figure we show the
{\em de-projected} circular velocities implied by our fit.  For stars
close to the line of nodes, the factor needed to deproject the
velocities becomes large, such that we have excluded stars with
projection factors larger than 5 from the left panel of Fig.\ 3 to
avoid confusion in the visual presentation; the \ion{H}{1} gas in the
region containing these excluded stars is shown as grayscale in the
right panel of Fig.\ 3.

As in Paper I, Fig.\ 3 shows that much of the \ion{H}{1} gas defines a
flat rotation curve; in contrast to Paper I, however, the much larger sample presented here yielded a new solution that
shows few significant RSG velocity outliers compared to the
\ion{H}{1}, and defines what appears to be a nearly identical rotation
curve to the \ion{H}{1}.  We also identify the velocity signatures of
the \ion{H}{1} arms S, E, and B \citep{ss03} and E2 (Paper I), and
their spatial location in the right-hand panel of Fig.\ 3.  We confirm
the Paper I result that few if any RSGs are associated with the
\ion{H}{1} arms, with a larger sample of RSG velocities.

In Fig.\ 4 (left), we compare the implied in-plane circular velocities
of our entire sample, which is predominantly made up of carbon-rich
and oxygen-rich AGB stars, with the kinematic model fit to the
RSG-only sample.  The bulk of the stars trace the envelope of the flat
rotation curve, but with many outliers with lower implied circular
velocities, including many negative circular velocities.  Examining
the line-of-sight velocity residuals of the full sample about the fit
to the RSGs (Fig.\ 4 right panel) reveals an offset in the peak of the
distribution, $<v_{\rm los,full}-v_{\rm los,RSG}>=-4.2$ km s$^{-1}$,
and a dispersion of 26 km s$^{-1}$.  The offset can be explained as
asymmetric drift, with the magnitude consistent with that expected
from the observed velocity dispersion, following vdM02.  In Paper I,
we found substantially different rotation velocities for LMC RSGs,
\ion{H}{1}, and carbon stars.  With the much larger sample of stars
studied here, it appears that these differences are only as large as
expected by asymmetric drift, such that all populations have basic
kinematics that are consistent with each other.

To understand the outliers better, we selected all stars with $\Delta v_{\rm los}>50$ km s$^{-1}$ (which have velocities in excess of those predicted by our kinematic model) and plotted them as red points in the left panel of Fig.\ 4, as well as all stars with  $\Delta v_{\rm los}<-50$ km s$^{-1}$, which are plotted as blue points.  Strikingly, the majority of these outliers fall in the region where we see the kinematic signatures of the E \ion{H}{1} arm, and its E2 extension, and the B arm.  In Paper I, we observed carbon stars in these regions of the plot, and interpreted them as stars that were being tidally stripped from the LMC along with the \ion{H}{1} gas; indeed, many of the stars appeared co-located with the \ion{H}{1} arms.  With our larger dataset, we see many more of these stars (376), as well as roughly equal numbers with positive ($\sim$40\%) and negative ($\sim$60\%) velocity residuals.
Fig.\ 5 shows the location of the stars with $|\Delta v_{\rm los}|>50$ km s$^{-1}$ on the sky, where the colors indicate the line-of-sight velocities after correction for the LMC's space motion and for d$i$/d$t$.  The underlying rotation of the LMC is visible in the bulk of the stellar population; it is worth noting that the near side of the LMC lies in the northeast (Caldwell \& Coulson 1986), such that the sense of rotation is clockwise on the page.  The stars with $|\Delta v_{\rm los}|>50$ km s$^{-1}$, however, not only have large residuals from the model, but clearly have line-of-sight velocities that oppose the sense of rotation of the disk.  Almost all of them also lie inside the outermost radius at which the LMC rotation curve can be traced.  If these stars lie in the LMC plane, then they must indeed be counter-rotating compared to the LMC.  If they lie in a structure with a very different geometry, e.g.\ in a plane whose near side is in the southwest rather than the northeast, then their sense of rotation could be the same as the LMC disk.
In either case, we think it is untenable to maintain the interpretation that these kinematically distinct stars have been tidally stripped from the LMC, as we concluded in Paper I based on a smaller sample of stars and on our expectation that the \ion{H}{1} arms with which the stars were associated were being pulled out of the LMC.  
We note that the fraction of this population, $\gtrsim$5\%, is similar to the 7\% of carbon stars that Graff et al.\ (2000) found comprised a kinematically distinct population in their analysis.  The similar fractions may be fortuitous, however, as we see no clear sign of the kinematically distinct population in the raw histogram of heliocentric velocities, contrary to what was seen by Graff et al.\ with their smaller sample.

To examine the geometry of the kinematically distinct population, we made the assumption that the stars are on circular planar orbits in the LMC's potential, for which we know the rotation curve from the RSGs above.  Fixing $v_{\rm sys}$, $v_{0}$ , and $R_{0}$ at the values determined for the LMC RSGs, we fit for the rate of change of inclination $($d$i$/d$t)_{\rm KDP}$, the angle of the kinematic line of nodes $\theta_{\rm KDP}$, and inclination $i_{\rm KDP}$ of the kinematically distinct population.  We found two solutions, the first being a counter-rotating solution with $($d$i$/d$t)_{\rm KDP}=290\arcdeg\pm50\arcdeg$, $\theta_{\rm KDP}=177\pm7\arcdeg$, and $i_{\rm KDP}=20\pm3\arcdeg$.  This solution would indicate that the kinematically distinct population lies in a plane that is twisted by $\sim35\arcdeg$ compared to the LMC disk and with an inclination that is $\sim15\arcdeg$ shallower with respect to the line of sight, but is qualititatively speaking roughly coplanar with the LMC.  This solution is similar to that adopted by Subramaniam \& Prabhu (2005), who modeled the LMC core as two disks with identical inclination but lines of nodes twisted by 40$\arcdeg$ from each other.
The second solution has the stars rotating in a clockwise sense on the sky, but with $($d$i$/d$t)_{\rm KDP}=300\arcdeg\pm50\arcdeg$, $\theta_{\rm KDP}=175\pm7\arcdeg$, and $i_{\rm KDP}=-19\pm2\arcdeg$.  In this solution, the kinematically distinct stars lie in a plane that is inclined by $\sim55\arcdeg$ compared to the LMC disk, similar to the fragmented polar ring model of Kunkel et al.\ (1997).  Neither solution depends strongly on the fitted value of (d$i$/d$t)_{\rm KDP}$; fixing (d$i$/d$t)_{\rm KDP}$ equal to the value derived for the LMC RSGs led to solutions with nearly identical $i_{\rm KDP}$, but with a more highly twisted line of nodes, $\theta_{\rm KDP}= 217\arcdeg\pm7\arcdeg$.  Both solutions have line-of-sight velocity dispersions about the solutions of 15 km s$^{-1}$, which is significantly smaller than the raw observed velocity dispersion of 26 km s$^{-1}$ for the kinematically distinct population.  The small velocity dispersions about the solutions implies $v/\sigma\sim$5 for the kinematically distinct population, demonstrating that the population is kinematically cold.

We next gathered evidence to understand the origin of the kinematically distinct population.  Fig.\ 6a shows the $J-[8.0], [3.6]$ CMD for the kinematically distinct population compared to all of the available sources.  The figure shows that nearly all of the kinematically distinct stars are O-rich and C-rich AGB stars, but are otherwise not remarkably different from the normal LMC AGB population, suggesting that they are at the same distance as the LMC.  The only difference appears to be that the normal LMC population contains some luminous O-rich and C-rich AGB stars that are not seen in the kinematically distinct population. To quantify this difference, we counted the number of LMC AGB stars found within the red outline in Fig. 6a, which are brighter than those in the polygon that contains the majority of the kinematically distinct population.  If the kinematically distinct population were drawn from the normal LMC AGB population, we would expect to observe 10$\pm$3 AGB stars brighter than those in the polygon, whereas we observe one star at the edge of the outline, with [J-8.0]$\sim$2.5 and [3.6]$\sim$7.  This star could be a bona fide bright AGB star or a red LMC RSG with a peculiar velocity.  A much more striking difference, shown in Fig.\ 6b, is seen in the spatial distribution of the kinematically distinct population compared to the normal LMC AGB population.  While normal LMC AGB stars are found at all locations within the SAGE survey, they are most heavily concentrated in the LMC Bar.  The kinematically distinct population, however, clearly avoids the Bar, implying a different origin from that of the normal LMC AGB stars.

We used our spectra to measure metallicities from the Ca II near-infrared triplet lines for a subset of stars in the kinematically distinct population and in the normal LMC population.  We used only those spectra without carbon star features and with S/N per resolution element in excess of $\sim$20, which yielded measurements for 994 normal LMC stars and 30 stars in the kinematically distinct population.  We followed Cole et al.\ (2004) to fit summed Gaussian and Lorentzian profiles to the spectral lines, integrated these fits to derive their equivalent widths, and formed Ca indices by summing the equivalent widths of the triplet lines.
After using the optical photometry and extinction map of Zaritsky et al.\ (2004) to correct the $V$ magnitudes of our stars for extinction, we used the calibration of Cole et al.\ (2004) to derive metallicities from the Ca indices and $V-V_{\rm HB}$, the magnitude differences between our stars and the LMC horizontal branch.  Following Cole et al.\ (2005), we adopted $V_{\rm HB}=19.22$ for all of the stars.  Note that although most of our stars are brighter than the globular cluster stars used to calibrate the Ca index/$V-V_{\rm HB}$/[Fe/H] relationship, we found that the normal LMC population defines the same slope in the Ca index/$V-V_{\rm HB}$ diagram as found by Cole et al.\ (2004), suggesting that the same calibration applies.  For the 994 normal LMC stars, we found a metallicity distribution with a peak at [Fe/H]=$-0.45$, a median of [Fe/H]=$-0.56\pm0.02$, and a dispersion of 0.5 dex compared to a median error of 0.15 dex.  These measurements are in good agreement with the LMC Bar metallicity distribution measured by Cole et al.\ (2005), who found a median [Fe/H]=$-0.4$ with a dispersion of 0.3 dex.  For the 30 stars from the kinematically distinct population for which we could measure the Ca triplet lines, we instead found a median [Fe/H] of $-1.25\pm0.13$ with dispersion 0.7 dex compared to a median error of 0.17 dex.  

\section{Discussion and Conclusions}

Where did the kinematically distinct population originate?  The stars have a kinematic signature that links them to \ion{H}{1} arms E and B.  \citet{ss03} identify these arms as connecting to the Leading Arm (Putman et al.\ 1998) in the case of arm E and the Magellanic Bridge and the SMC in the case of arm B.  The most tempting and obvious answer is thus that the stars originated in the SMC, and have fallen into the LMC as a result of the LMC-SMC-Milky Way interaction.  However, both arms E and B are thought to be outflows, perhaps driven by a combination of star formation activity and tidal forces (Nidever et al.\ 2008), whereas an association with stars captured from the SMC would appear to require that they represent gas {\em infall}.  We cannot discount the possibility of infall, as without knowledge of the gas geometry, its direction of motion is inherently ambiguous.  An infall hypothesis for the arm E and B gas would indeed yield a natural explanation for two puzzling aspects of the LMC.  First, \ion{H}{1} maps show that the southeastern rim of the LMC, near the location of 30 Dor, contains an overdensity of \ion{H}{1} gas (e.g.\ Kim et al.\ 1998), and a similar overabundance of CO and giant molecular clouds (Fukui et al.\ 1999).  The morphology of the gas led de Boer et al.\ (1998) to conclude that the overabundance exists because gas piles up on the leading edge of the LMC due to ram pressure as the LMC moves through the Galactic halo, while Nidever et al.\ (2008) identify the overdensity as the source of the outflows which they concluded feed arms E and B.  In the infall hypothesis, the overdensity would instead be produced as the infalling gas encounters the LMC disk, either in a nearly head-on collision (if the kinematically distinct population is counter-rotating) or in a ``T-bone'' collision (if the geometry of the kinematically distinct population is highly inclined with respect to the LMC).  The stars would continue through the site of the collision unimpeded, while the gas would shock as it encountered the LMC gaseous disk.  The infall hypothesis could thus account for a second puzzling aspect of the LMC, namely the unique properties of the 30 Dor star forming complex.  As discovered by Dolphin \& Hunter (1998), the region of Constellation III in the LMC formed from roughly the same gas mass as 30 Dor, but 30 Dor has yielded several times more stellar mass and at higher spatial concentration, including the dense central cluster like R136.  Dolphin \& Hunter suggested that this was because the initial distribution of gas in 30 Dor was much more compact, something which a violent collision of gas would naturally explain.

Our measurements of the Ca triplet abundances provide a check on the hypothesis that the kinematically distinct stars came from the SMC.  De Propris et al.\ (2010) used the Ca triplet to measure abundances of hundreds of SMC red giants at the periphery of the SMC, and found a metallicity distribution with mean [Fe/H]=$-1.35\pm0.10$ and a dispersion of 0.65$\pm$0.08 dex, in very good agreement with our measurement of the metallicity distribution of the kinematically distinct population.  Although we would like to confirm this result with high signal-to-noise abundance measurements of many more stars, the Ca triplet abundances provide strong evidence that the kinematically distinct population did originate in the SMC.  We find further evidence that the stars originated in the SMC from Fig.\ 7, which shows a comparison of the $J-[3.6]$,$[3.6]$ CMD of the SMC (Gordon et al., submitted) with the CMD of the kinematically distinct stars as shifted to the distance of the SMC.  In making this figure, we have assumed that the SMC has distance modulus 0.4 mags larger than the LMC, which is based on an LMC distance modulus of 18.5 and an SMC modulus of 18.9 (Storm et al.\ 2004).  We have ignored extinction by dust, since for our stars we found $<A_V>=0.4$, which translates to negligibly small $A_J\sim0.1$ (Schlegel et al.\ 1998) and $A_{[3.6]}=0.02$ (Indebetouw et al.\ 2005).  The figure demonstrates that the CMDs of the SMC and kinematically distinct LMC stars are qualitatively similar, particularly if we restrict the SMC CMD to include only stars beyond radii of $\sim$2 degrees from the optical center of the SMC, which we take to be $\alpha=0^{\rm h}49^{\rm m}47^{\rm s}$, $\delta=-72\arcdeg53\arcmin40\arcsec$ (Westerlund 1997).  At these outer radii, the most luminous AGB stars that are seen in the full SMC CMD, but not in the kinematically distinct population, no longer appear.  To quantify this distinction, we counted the number of SMC AGB stars found within the area defined by the red outline in Fig. 6a (shifted vertically by +0.4 magnitudes to account for the difference in distance modulus between the SMC and LMC) and compared it to the number of stars with properties similar to the kinematically distinct population, as defined by the polygon in Fig.\ 6a (again shifted for distance modulus).  If the kinematically distinct population were drawn from the full SMC AGB population, we found that we would expect to observe 6$\pm$2 bright AGB stars in the sample, whereas we observe one.  If the kinematically distinct population were instead drawn from the SMC AGB population beyond a 2$\arcdeg$ radius from the center, we would expect to see 1$\pm$1 bright AGB star, and we observe one.  Although these values are subject to small number statistics, such agreement with the stellar populations in the outer regions of the SMC is just what we would expect if the kinematically distinct stars were stripped from the SMC in a close passage between the Clouds.

Coupled with our comparison of the stellar and \ion{H}{1} kinematics, our results require a reinterpretation of the formation of \ion{H}{1} arms E and B, and suggest that we should also examine our findings in light of models designed to form the Leading Arm feature and the Magellanic Stream.  A recent model is that of Besla et al.\ (2010), who have successfully reproduced the basic properties of the Stream and the Leading Arm.  In their model, the Clouds are a binary system that experienced a close interaction 1.2 Gyr ago, during which the Stream was formed from gas tidally removed from the SMC by the LMC.  The SMC is modeled as having a compact stellar disk and an extended gas disk, resulting in no stripped SMC stars, and the Stream without an observable stellar component.  Our result argues, however, that $\sim5\%$ of the stars in the LMC came from the SMC, such that the Besla et al. model needs some modification.  Indeed, with a more extended SMC stellar disk, the Besla et al. model {\em predicts} that SMC stars should be captured by the LMC (Besla, personal communication); investigating the geometry and kinematic properties of these captured stars requires more modeling work.  The existence of accreted SMC stars in the LMC would then also predict that SMC stars should be present in the Stream.  Their surface brightness would still be predicted to be lower than current detection limits, but should be reachable by deep wide-field imaging with current or future facilities.

In summary, we have used $\sim$4600 new spectra and $\sim$1300 published velocities of primarily red supergiants and AGB stars to study the internal kinematics of the LMC.  Our main conclusions are:

\begin{enumerate}
\item After correcting the line-of-sight velocities for the LMC's space motion and accounting for asymmetric drift in the AGB population, we have found a rotation curve that is consistent with the red supergiants, the \ion{H}{1}, and the AGB stars.
\item We have discovered a population of stars, representing $\sim$5\% of our sample, that has distinct kinematics.  
\item The kinematically distinct population is comprised almost entirely of AGB stars, but do not have the number of bright AGB stars that would be expected if they were drawn from the normal LMC AGB population.  Moreover, they have a spatial distribution that avoids the LMC Bar, contrary to the normal LMC AGB population.
\item Our simple calculation finds that the kinematically distinct stars are either counter-rotating in a plane closely aligned with the LMC disk, or rotating in the same sense as the LMC disk, but in a plane that is inclined by 54$\arcdeg\pm2\arcdeg$ to the LMC.  Their kinematics clearly link them to two known \ion{H}{1} arms, which have previously been interpreted as being pulled out from the LMC.  
\item The metallicities of the kinematically distinct population, measured from the Ca triplet lines, are low compared to the LMC but in good agreement with the metallicities of SMC giants.
\item The $J-[3.6], [3.6]$ color-magnitude diagram of the kinematically distinct population is a good visual match to that of the outer regions of the SMC when shifted to the same distance.
\item Our conclusion that the kinematically distinct stars found in the LMC were accreted from the SMC implies that the associated \ion{H}{1} arms, previously thought to be flowing away from the LMC, are likely to be falling in.  This interpretation could help to explain the overdensity of gas seen on the LMC's southwestern rim, and identifies a possible source of fuel for star formation in 30 Doradus.
\end{enumerate}

\acknowledgements
We would like to thank Dara Norman for suggestions throughout this project, Phil Massey for a careful reading of this paper and for helpful comments, and Gurtina Besla for providing valuable insights. We also thank Verne Smith, Nathan Smith, and Simon Schuler for useful suggestions.  We thank Eduardo Hardy, David Graff, and Nick Suntzeff for providing their carbon star data, and the CTIO mountain support staff for making these observations possible.  Finally, we thank an anonymous referee for helpful comments that improved this paper.

\clearpage
\begin{figure}
\plotone{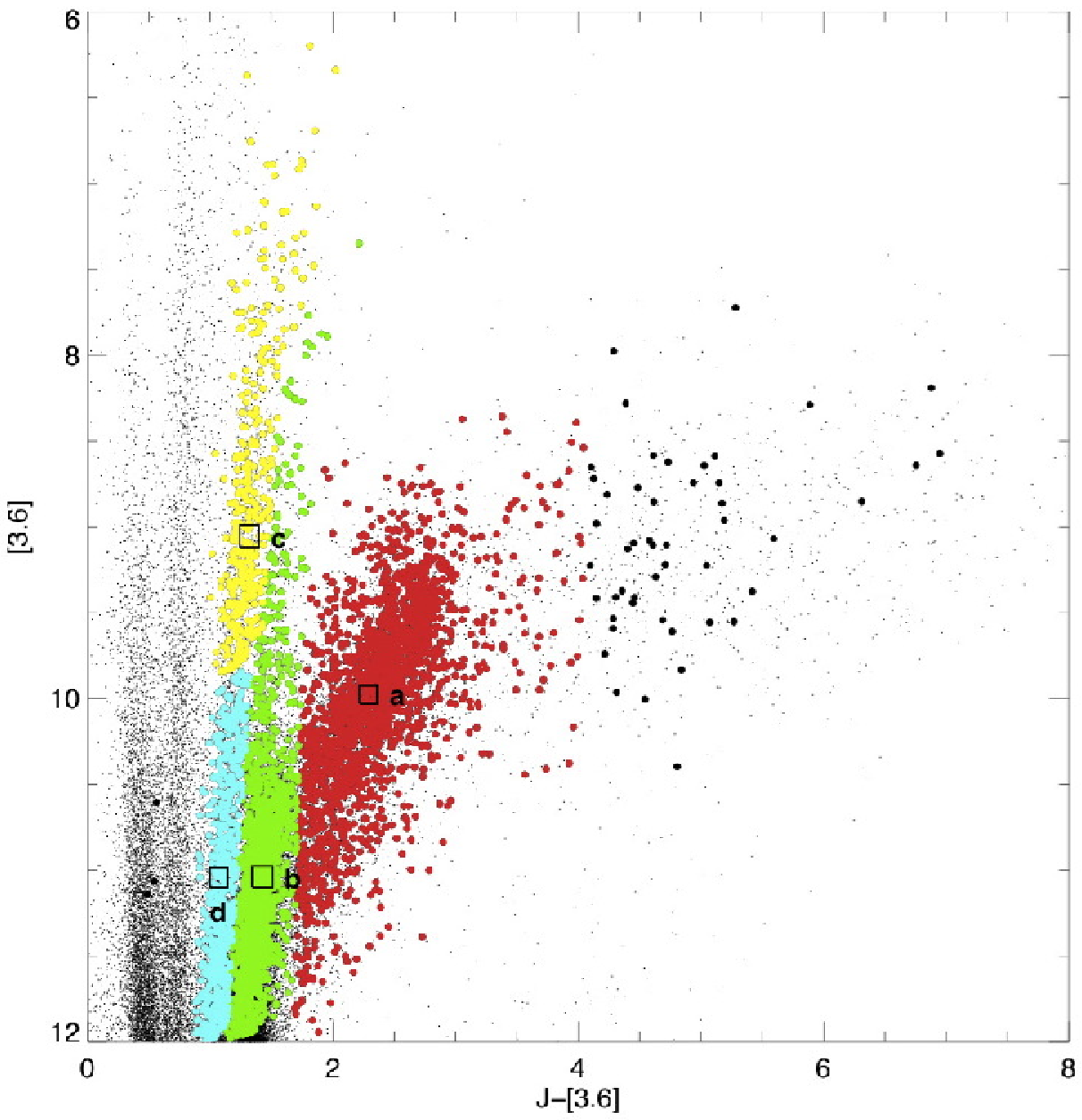}
\caption{The color-magnitude diagram of all sources toward the LMC measured by SAGE (black points; Meixner et al.\ 2006).  Filled circles are those stars for which we obtained spectra.  The colors of the filled circles denote our rough classification of the stars, based on the criteria established by Blum et al.\ (2006): yellow for red supergiants, cyan for giants, green for oxygen-rich AGB stars, red for carbon stars, and black for ``extreme'' AGB stars.  Boxes identify areas within which we randomly selected a star to represent its class.}
\end{figure}
\renewcommand{\thefigure}{\arabic{figure}\alph{subfigure}}
\setcounter{subfigure}{1}
\begin{figure}
\epsscale{1.0}
\plotone{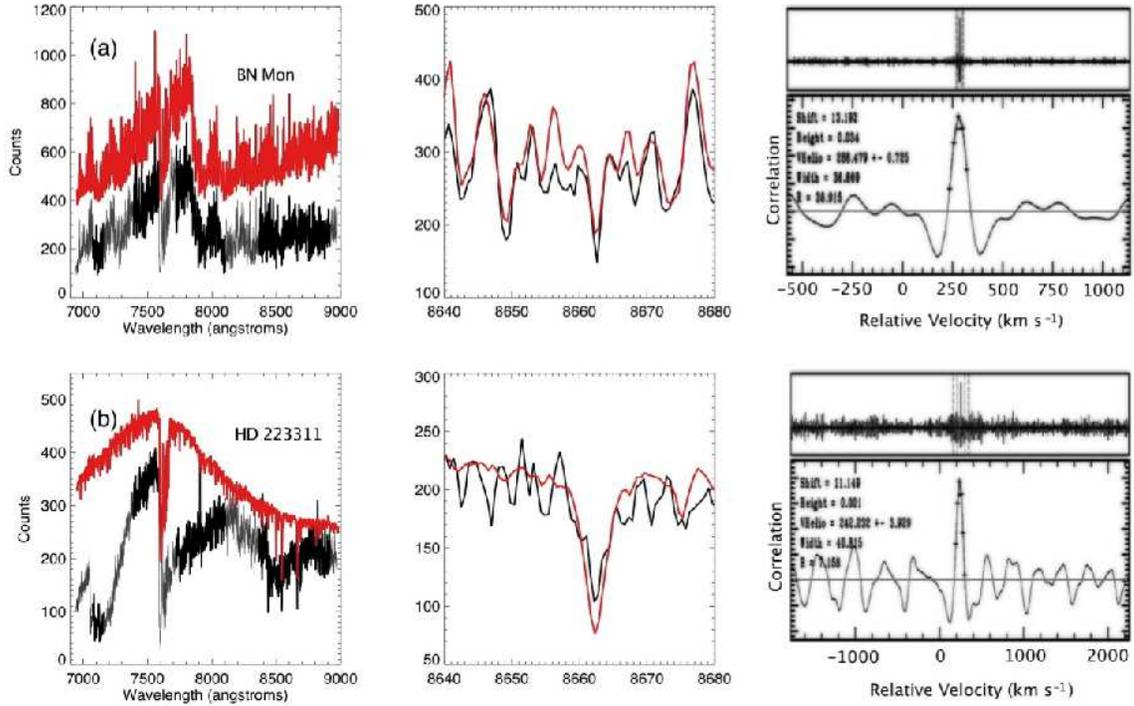}
\caption{Spectra of a representative LMC carbon star (a) and oxygen-rich AGB star (b) are shown, along with the velocity template used for each star.  {\it Left column:} The LMC spectra are shown in black, with grey marking wavelengths contaminated by terrestrial absorption.  The spectra have been shifted in wavelength according to the velocity computed by cross-correlation with the template.  The template spectra are shown in red, are labelled with names, and have been shifted vertically to avoid confusion.  {\it Middle column:}The LMC spectra are overlaid with the templates in a narrow wavelength region centered on the 8662 \AA ~Ca II triplet line.  {\it Right column:} The correlation strength computed by FXCOR is shown as a function of the relative velocity shift.  The upper panels show the correlations over their full range, while the lower panels present a zoomed-in view of the peaks at the computed relative shift.  The grey horizontal lines mark the zero levels.}
\end{figure}
\addtocounter{figure}{-1}
\addtocounter{subfigure}{1}
\begin{figure}
\epsscale{1.0}
\plotone{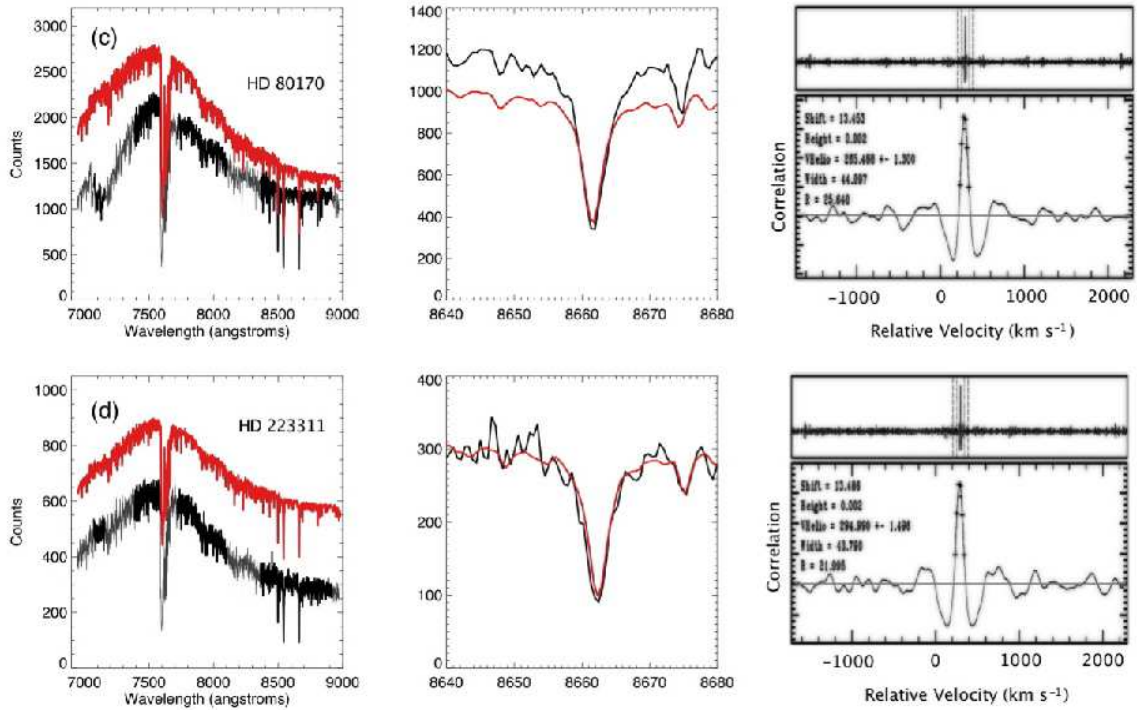}
\caption{Spectra of a representative LMC red supergiant (c) and $\sim$5 M$_\odot$ giant (d) are shown, along with the velocity template used for each star.  See Fig.\ 2a for a description of the plots.}
\end{figure}
\renewcommand{\thefigure}{\arabic{figure}}
\begin{figure}
\epsscale{1.0}
\plotone{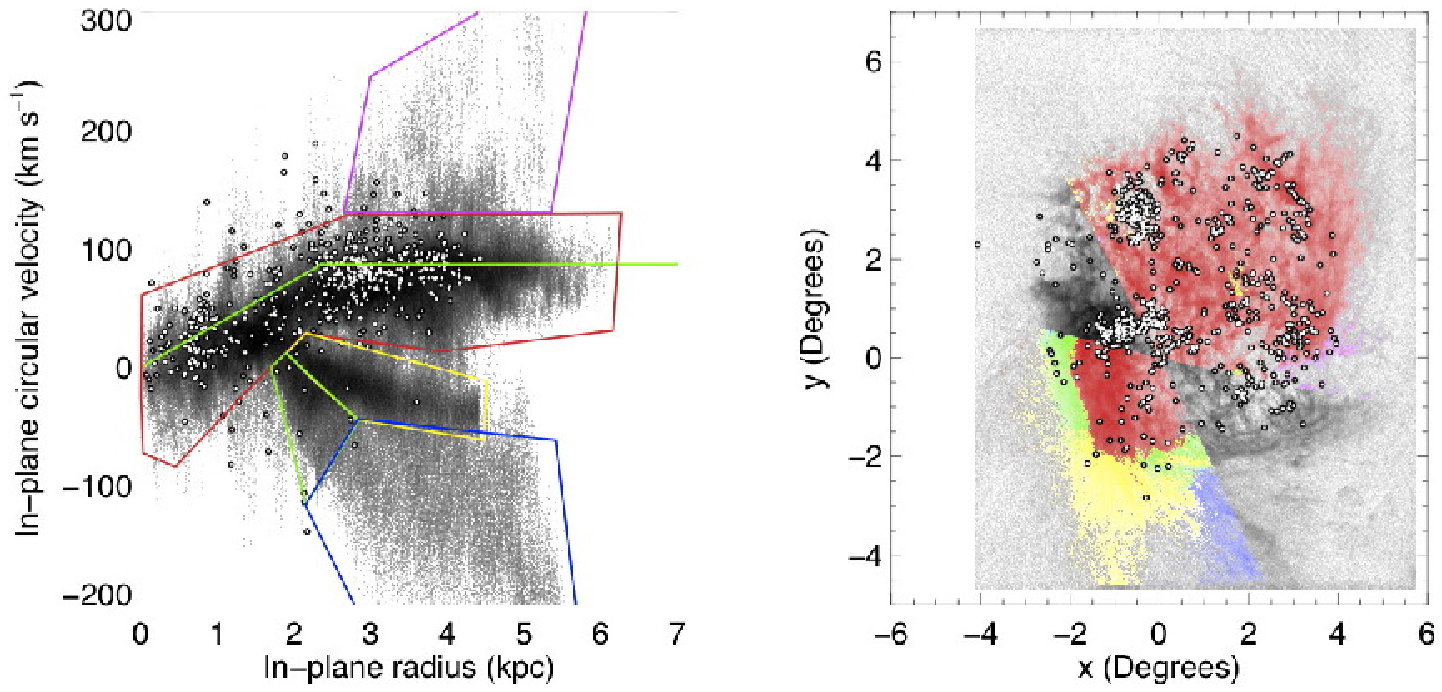}
\caption{The rotation curve of the LMC as fit to red supergiants.  {\it Left:} The line-of-sight velocities of red supergiants (circles) and of \ion{H}{1} gas (grayscale; Kim et al.\ 1998) have been deprojected to show their in-plane circular velocities, as described in the text.  The green line is our parameterized fit of the rotation curve to the RSGs.  The colored polygons identify the distinct kinematic components discussed in paper I, most of which were identified by Staveley-Smith et al.\ (2003): the flat rotation curve (red), arm S (purple), arm E and its E2 extension (yellow and green), and arm B (blue).  {\it Right:} The \ion{H}{1} gas and red supergiants are shown as projected on the sky, with color-coding corresponding to the kinematic components identified at left.}
\end{figure}
\begin{figure}
\plotone{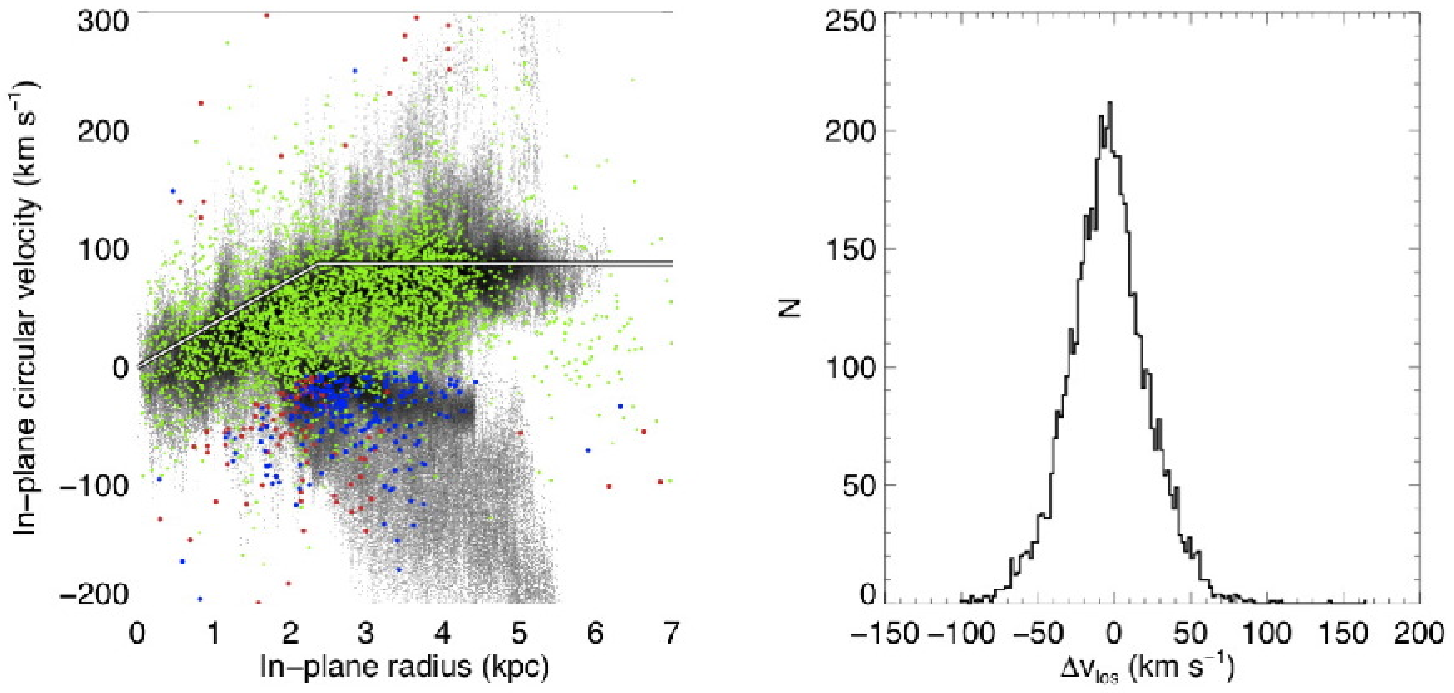}
\caption{The kinematics of all stars studied in this work.  On the left, the red supergiant rotation curve (white line) is compared to the deprojected in-plane circular velocities of all stars studied herein; these stars are mostly carbon-rich and oxygen-rich AGB stars (points).  The red and blue points have line-of-sight velocity residuals, whose distribution is shown on the right, that exceed 50 km s$^{-1}$.  These stars represent the tails of the distribution.}
\end{figure}
\begin{figure}
\plotone{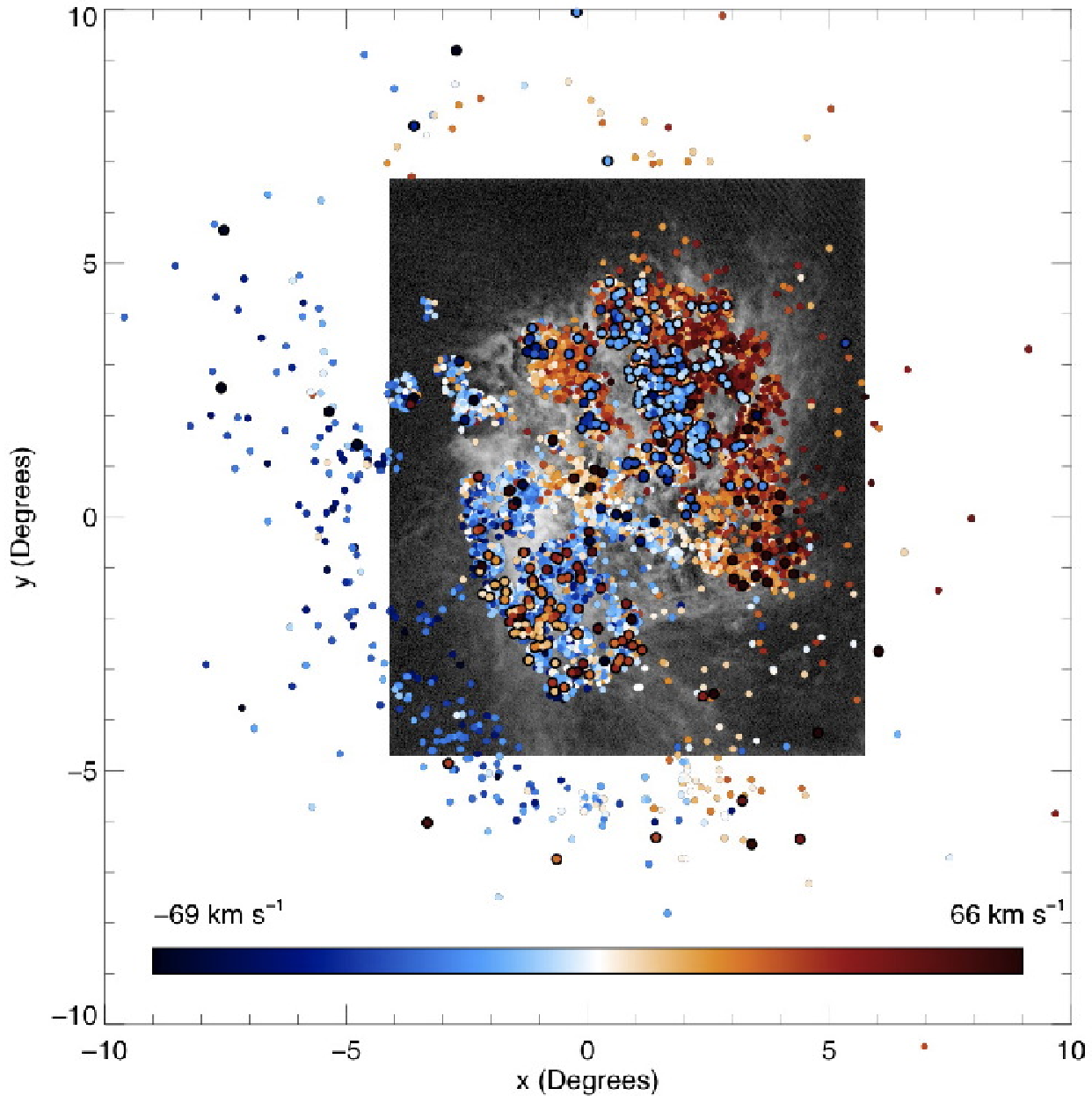}
\caption{The \ion{H}{1} gas (greyscale) and stars (points) are shown projected on the sky, where the color shading indicates the line-of-sight velocities after correction for the LMC's space motion.  The rotation of the LMC disk is seen as a gradient running from blue in lower left corner to red in the upper right.  $\sim$5\% of the stars, however, emphasized by a larger point size, have the opposite gradient.}
\end{figure}
\begin{figure}
\plottwo{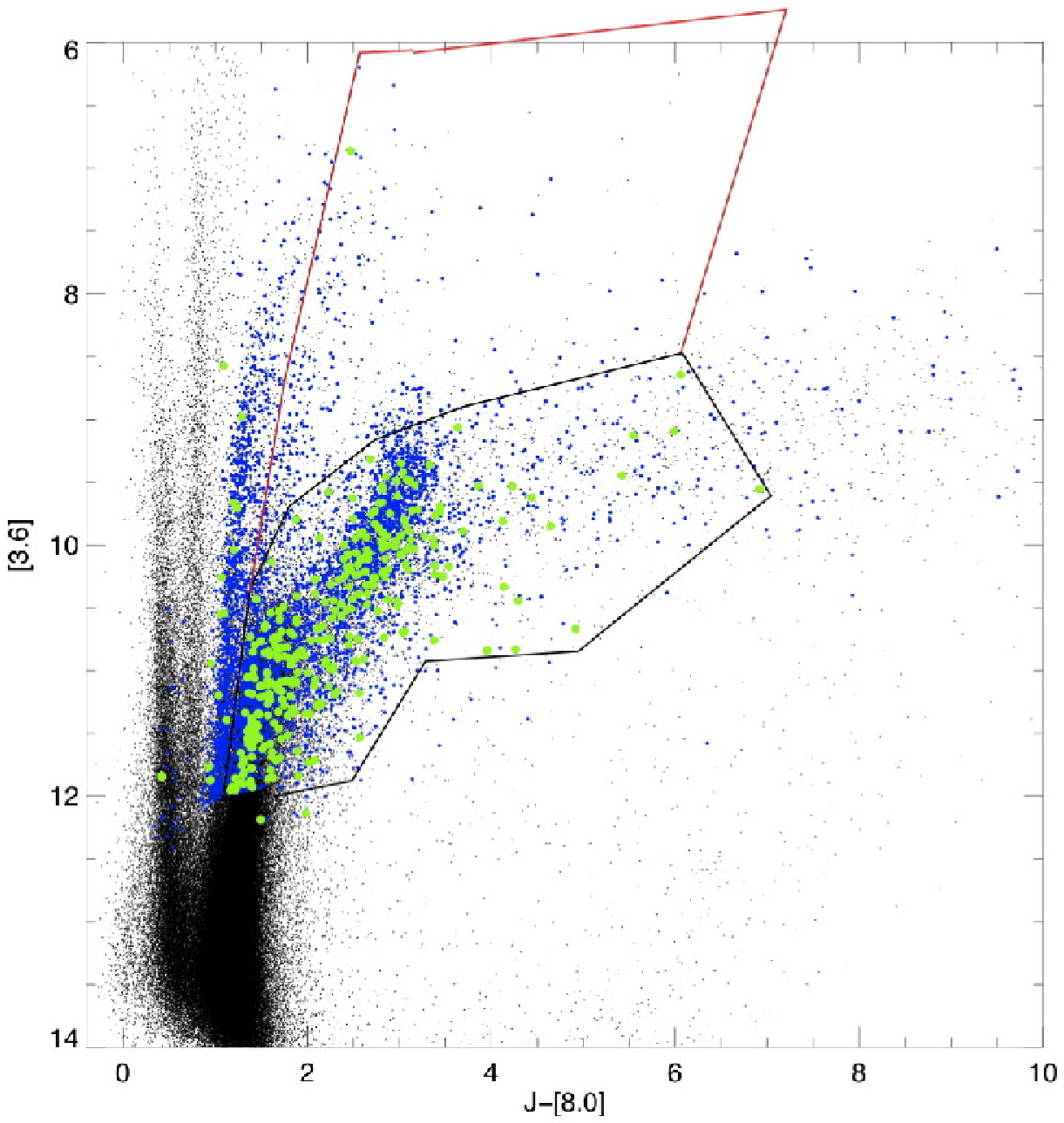}{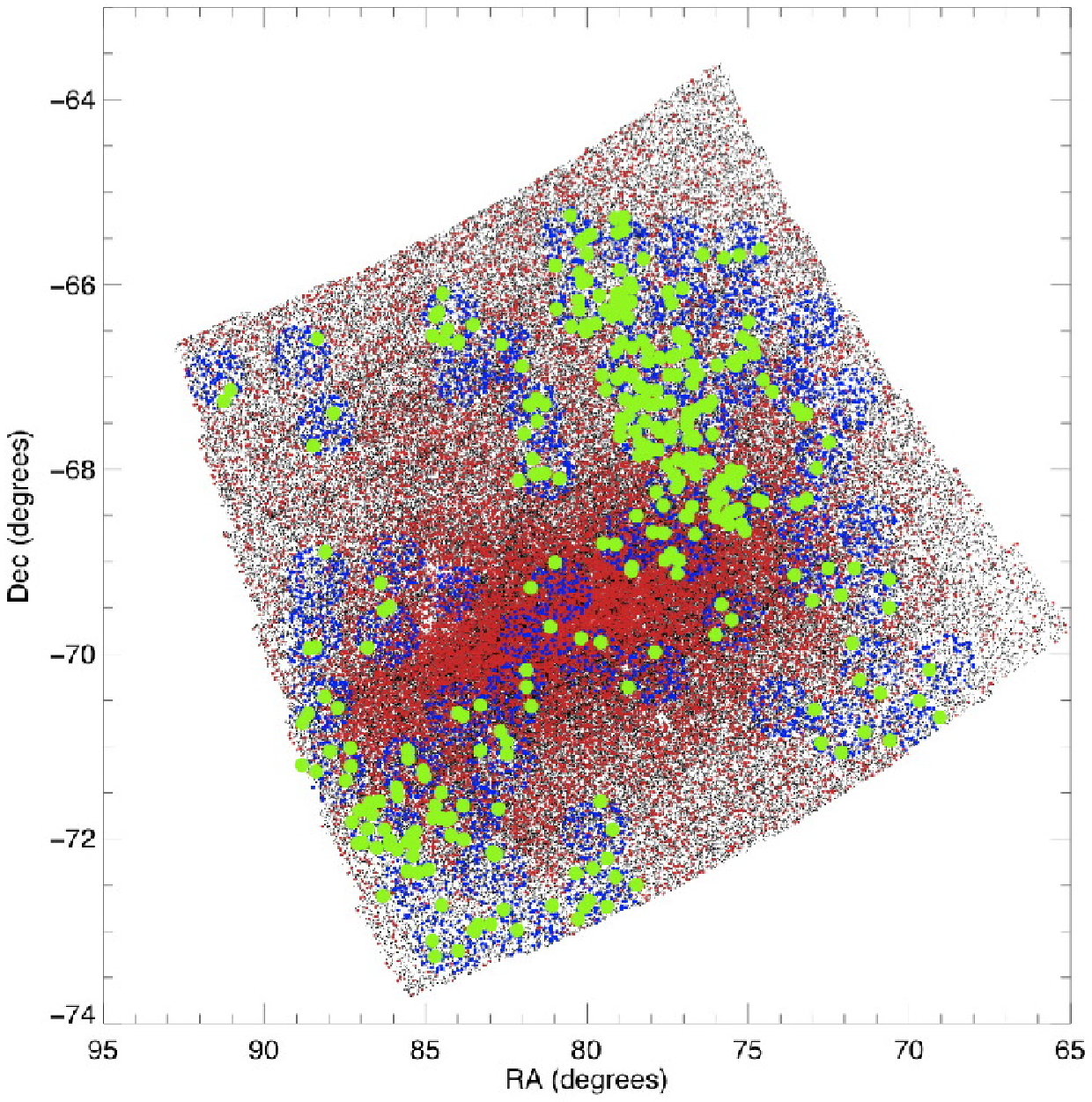}
\caption{{\it Left:} The color-magnitude diagram of all sources toward the LMC measured by SAGE (black points; Meixner et al.\ 2006).  Points in blue are LMC stars with measured optical spectra.  Green points are the stars with line-of-sight velocity residuals in excess of 50 km s$^{-1}$, which we identify as a kinematically distinct population in the LMC.  Almost all of these stars appear to be AGB stars, with properties not grossly different from the normal LMC AGB population, with the exception that the most luminous oxygen-rich and carbon-rich AGB stars are not represented in the kinematically distinct population.  {\it Right:} The spatial distribution of all SAGE LMC point sources is shown as black points.  Red points show LMC AGB stars as defined by those sources that fall within the polygon in the plot on the left; these stars concentrate towards the LMC Bar.  Blue circles show stars for which we obtained spectra, while green circles represent the kinematically distinct population.  Although our spectra are not uniformly sampled spatially, the kinematically distinct population clearly avoids the Bar, contrary to the normal LMC AGB population.}
\end{figure}
\begin{figure}
\epsscale{0.7}
\plotone{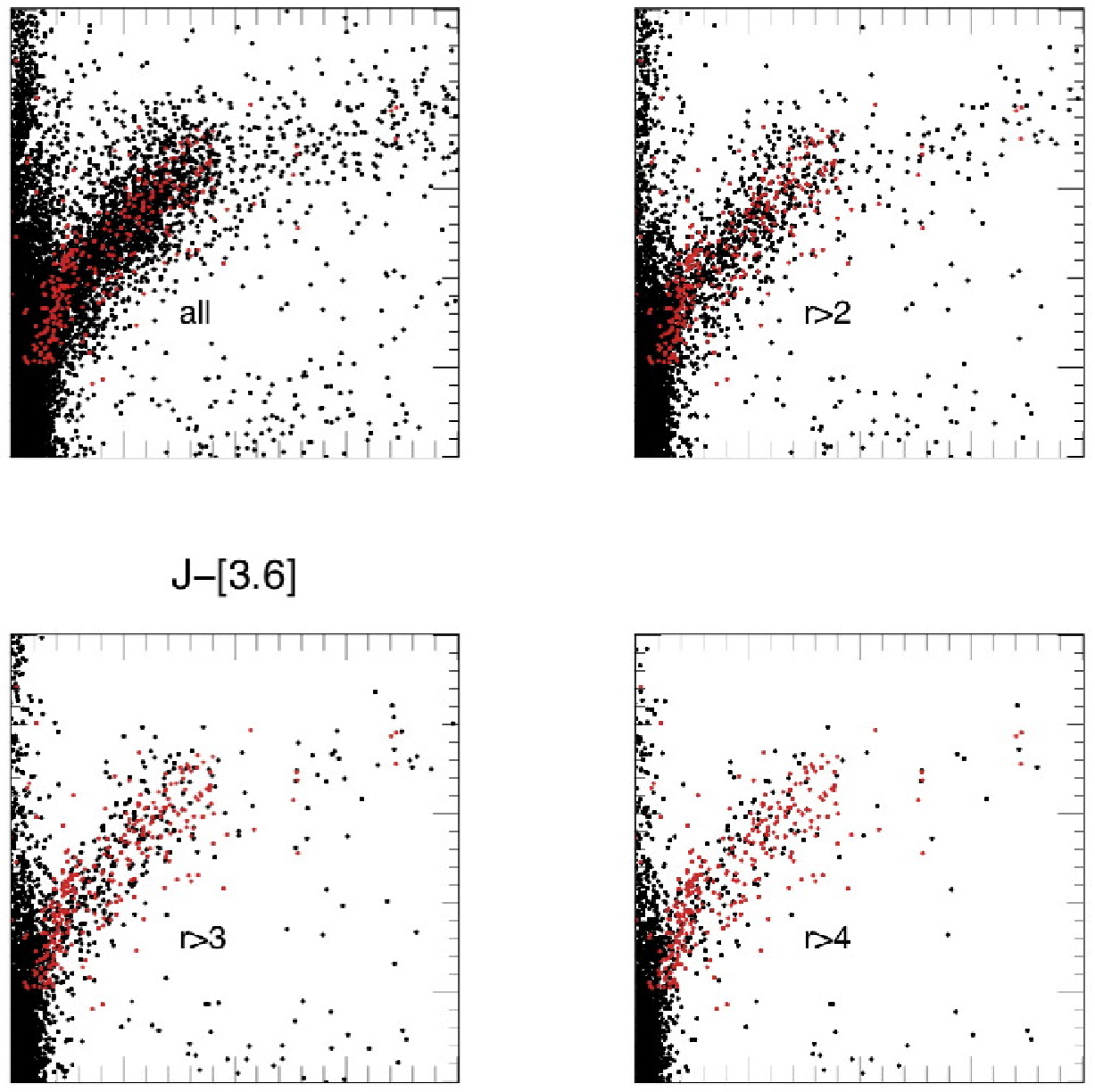}
\caption{The color-magnitude diagram of sources toward the SMC as measured by SAGE-SMC (black points; Gordon et al., submitted), plotted with four different radial cuts.  The top left panel shows the SMC CMD for all sources in SAGE-SMC, while the top right, bottom left, and bottom right panels only include sources beyond 2$\arcdeg$, 3$\arcdeg$, and 4$\arcdeg$ radius, respectively, from the optical center of the SMC.  The red points are the kinematically distinct stars seen in the LMC from this work, shifted by $+0.40$ magnitudes in [3.6] to account for the difference in distance modulus between the LMC and SMC.  While the CMD containing all SMC sources has many oxygen-rich and carbon-rich AGB stars that are more luminous than the ones we see in the kinematically distinct population, at larger radii these luminous AGB stars no longer appear.  There is, in particular, a good visual match between the SMC CMD at radii beyond 3$\arcdeg$ and the shifted CMD of the kinematically distinct population in the LMC.}
\end{figure}


\newpage
\begin{thebibliography}{}
\bibitem[Baranne et al.(1979)]{1979VA.....23..279B} Baranne, A., Mayor, M., 
\& Poncet, J.~L.\ 1979, Vistas in Astronomy, 23, 279 
\bibitem[Barden \& Ingerson(1998)]{barden98} Barden, S.~C.~\& 
Ingerson, T.~E.\ 1998, ASP Conf.~Ser.~152: Fiber Optics in Astronomy III, 
60 
\bibitem[Besla et al.(2007)]{besla07} Besla, G., Kallivayalil, N., Hernquist, L., Robertson, B., Cox, T.J., van der Marel, R.P., Alcock, C.\ 2007, \apj, 668, 949
\bibitem[Besla et al.(2010)]{besla10} Besla, G., Kallivayalil, N., Hernquist, L., van der Marel, R.P., Cox, T.J., Keres, D.\ 2010, \apjl, 721, L97
\bibitem[Blum et al.(2006)]{Blum06} Blum, R.D., et al.\ 2006, \aj, 132, 2034
\bibitem[Boylan-Kolchin et al.(2010)]{bk10} Boylan-Kolchin, M., Besla, G., \& Hernquist, L.\ 2010, arXiv:10104797
\bibitem[Caldwell \& Coulson(1986)]{cc86} Caldwell, J.~A.~R., \& Coulson, I.~M.\ 1986, \mnras, 218, 223 
\bibitem[Cole et al.(2004)]{cole04} Cole, A.~A., Smecker-Hane, 
T.~A., Tolstoy, E., Bosler, T.~L., \& Gallagher, J.~S.\ 2004, \mnras, 347, 367 
\bibitem[Cole et al.(2005)]{cole05} Cole, A.~A., Tolstoy, E., 
Gallagher, J.~S., III, \& Smecker-Hane, T.~A.\ 2005, \aj, 129, 1465 
\bibitem[de Boer et al.(1998)]{deboer98} de Boer, K.~S., Braun, J.~M., Vallenari, A., \& Mebold, U.\ 1998, \aap, 329, L49 
\bibitem[De Propris et al.(2010)]{depropris10} De Propris, R., 
Rich, R.~M., Mallery, R.~C., \& Howard, C.~D.\ 2010, \apjl, 714, L249
\bibitem[Dolphin \& Hunter(1998)]{dolphin98} Dolphin, A.~E., \& Hunter, D.~A.\ 1998, \aj, 116, 1275
\bibitem[Fukui et al.(1999)]{fukui99} Fukui, Y., et al.\ 1999, 
\pasj, 51, 745
\bibitem[Graff et al.(2000)]{2000ApJ...540..211G} Graff, D.~S., Gould, 
A.~P., Suntzeff, N.~B., Schommer, R.~A., \& Hardy, E.\ 2000, \apj, 540, 211 
\bibitem[Hardy et al.(2001)]{hardy} Hardy, E., Alves, D.~R., 
Graff, D.~S., Suntzeff, N.~B., 
\& Schommer, R.~A.\ 2001, Astrophysics and Space Science Supplement, 277, 471 
\bibitem[Harris(2007)]{harris07} Harris, J.\ 2007, \apj, 658, 
345 
\bibitem[Indebetouw et al.(2005)]{2005ApJ...619..931I} Indebetouw, R., et 
al.\ 2005, \apj, 619, 931 
\bibitem[Kallivayalil et al.(2006a)]{kallivayalil06a} Kallivayalil, N., 
van der Marel, R.~P., Alcock, C., Axelrod, T., Cook, K.~H., Drake, A.~J., 
\bibitem[Kallivayalil et al.(2006b)]{kallivayalil06b} Kallivayalil, N., 
van der Marel, R.~P., \& Alcock, C.\ 2006, \apj, 652, 1213
\bibitem[Kim et al.(1998)]{kim98} Kim, S., Staveley-Smith, 
L., Dopita, M.~A., Freeman, K.~C., Sault, R.~J., Kesteven, M.~J., \& 
McConnell, D.\ 1998, \apj, 503, 674 
\bibitem[Kunkel et al.(1997)]{kunkel97} Kunkel, W.~E., Demers, 
S., Irwin, M.~J., \& Albert, L.\ 1997, \apjl, 488, L129 
\bibitem[Massey(1995)]{massey95} Massey, P.\ 1995, hydraassign User Manual
\bibitem[Massey \& Olsen(2003)]{massey03} Massey, P., \& Olsen, 
K.~A.~G.\ 2003, \aj, 126, 2867 
\bibitem[Mathewson et al.(1974)]{mathewson74} Mathewson, D.~S., 
Cleary, M.~N., \& Murray, J.~D.\ 1974, \apj, 190, 291
\bibitem[Meixner et al.(2006)]{meixner06} Meixner, M., et al.\ 2006, \aj, 132, 2268
\bibitem[Mu\~{n}oz et al.(2006)]{munoz06} Mu{\~n}oz, R.~R., et 
al.\ 2006, \apj, 649, 201 
\bibitem[Nidever et al.(2008)]{nidever08} Nidever, D.~L., 
Majewski, S.~R., \& Burton, W.~B.\ 2008, \apj, 679, 432 
\bibitem[Nidever et al.(2010)]{nidever10} Nidever, D.~L., 
Majewski, S.~R., Butler Burton, W., \& Nigra, L.\ 2010, \apj, 723, 1618
\bibitem[Nikolaev et al.(2004)]{nikolaev04} Nikolaev, S., Drake, 
A.~J., Keller, S.~C., Cook, K.~H., Dalal, N., Griest, K., Welch, D.~L., 
\& Kanbur, S.~M.\ 2004, \apj, 601, 260 
\bibitem[Olsen \& Massey(2007)]{olsen07} Olsen, K.~A.~G., \& Massey, P.\ 2007, \apjl, 656, L61 (Paper I)
\bibitem[Olsen \& Salyk(2002)]{olsen02} Olsen, K.~A.~G., \& Salyk, C.\ 2002, \aj, 124, 2045
\bibitem[Piatek et al.(2008)]{piatek08} Piatek, S., Pryor, C., \& Olszewski, E.~W. 2008, \aj, 135, 1024
\bibitem[Pr\'{e}vot et al.(1985)]{prevot85} Prevot, L., et al.\ 
1985, \aaps, 62, 23 
\bibitem[Putman et al.(1998)]{putman98} Putman, M.~E., et al.\ 
1998, \nat, 394, 752 
\bibitem[Rix et al.(1992)]{rix92} Rix, H.-W., Franx, M., 
Fisher, D., \& Illingworth, G.\ 1992, \apjl, 400, L5 
\bibitem[Rubin et al.(1992)]{rubin92} Rubin, V.~C., Graham, 
J.~A., \& Kenney, J.~D.~P.\ 1992, \apjl, 394, L9 
\bibitem[Storm et al.(2004)]{2004A&A...415..531S} Storm, J., Carney, B.~W., Gieren, W.~P., Fouqu{\'e}, P., Latham, D.~W., \& Fry, A.~M.\ 2004, \aap, 415, 531 
\bibitem[Tonry \& Davis(1979)]{tonry79} Tonry, J.~\& Davis, M.\ 
1979, \aj, 84, 1511 
\bibitem[Saha et al.(2010)]{saha10} Saha, A., et al.\ 2010, 
\aj, 140, 1719 
\bibitem[Schlegel et al.(1998)]{1998ApJ...500..525S} Schlegel, D.~J., 
Finkbeiner, D.~P., \& Davis, M.\ 1998, \apj, 500, 525 
\bibitem[Silva \& Cornell(1992)]{1992ApJS...81..865S} Silva, D.~R., \& Cornell, M.~E.\ 1992, \apjs, 81, 865 
\bibitem[Skrutskie et al.(2006)]{skrutskie06} Skrutskie, M.F. et al.\ 2006, \aj, 131, 1163
\bibitem[Staveley-Smith et al.(2003)]{ss03} Staveley-Smith, 
L., Kim, S., Calabretta, M.~R., Haynes, R.~F., \& Kesteven, M.~J.\ 2003, \mnras, 339, 87 
\bibitem[Subramaniam \& Prabhu(2005)]{sp05} Subramaniam, A., \& Prabhu, T.~P.\ 2005, \apjl, 625, L47
\bibitem[Subramaniam \& Subramanian(2009)]{ss09} Subramaniam, A., \& Subramanian, S.\ 2009, \apjl, 703, L37 
\bibitem[van der Marel(2001)]{vdM01} van der Marel, R.~P.\ 
2001, \aj, 122, 1827 
\bibitem[van der Marel \& Cioni(2001)]{vdMC01} van der Marel, R.~P., \& Cioni, M.-R.~L.\ 2001, \aj, 122, 1807
\bibitem[van der Marel et al.(2002)]{vandermarel02} van der Marel, R.~P., Alves, D.~R., Hardy, E., \& Suntzeff, N.~B.\ 2002, \aj, 124, 2639 (vdM02)
\bibitem[van Dokkum(2001)]{vandokkum01} van Dokkum, P.~G.\ 2001, \pasp, 113, 1420 
\bibitem[Vergani et al.(2007)]{vergani07} Vergani, D., Pizzella, A., Corsini, E.~M., van Driel, W., Buson, L.~M., Dettmar, R.-J., \& Bertola, F.\ 2007, \aap, 463, 883
\bibitem[Wannier \& Wrixon(1972)]{ww72} Wannier, P., \& Wrixon, G.~T.\ 1972, \apjl, 173, L119 
\bibitem[Westerlund(1997)]{westerlund97} Westerlund, B.~E.\ 1997, The Magellanic Clouds (Cambridge: Cambridge Univ.\ Press)
\bibitem[Zaritsky et al.(2004)]{zaritsky04} Zaritsky, D., Harris, 
J., Thompson, I.~B., \& Grebel, E.~K.\ 2004, \aj, 128, 1606

\end{thebibliography}
\end{document}